\documentclass[11pt,twoside]{article}
\usepackage{cozumel2005}
\usepackage{epsf}
\usepackage{psfig}
\usepackage{lscape}
\usepackage{graphicx}
\begin{document}
\title{Variable stars in the field and in the clusters of the Fornax dwarf spheroidal galaxy}    
\author{C., Greco$^{1,2}$, G., Clementini$^1$, E. V., Held$^3$, E., Poretti$^4$, M., Catelan$^5$,  L., Dell'Arciprete$^4$, M., Gullieuszik$^{3,6}$, M.,  Maio$^2$, L.
Rizzi$^7$, H. A., Smith$^8$, B. J., Pritzl$^9$,  A., Rest$^{10}$, N., De Lee$^8$}   
\affil{$^1$ INAF- Osservatorio Astronomico di Bologna, Via Ranzani, 1, 40127, Bologna, Italy\\
{$^2$ Dipartimento di Astronomia, Universit\`a di Bologna, Via Ranzani, 1, 40127, Bologna, Italy\\
 $^3$ INAF- Osservatorio Astronomico di Padova, Vicolo dell'Osservatorio 5, 35122, Padova, Italy\\ 
 $^4$ INAF- Osservatorio Astronomico di Brera, Via E. Bianchi 46, 23807 Merate, Italy\\
  $^5$ Departamento de Astronom\'ia y Astrof\'isica, Pontificia Universidad Cat\'olica de Chile, Avenida Vicu\~na Mackenna 4860,
782-0436 Macul, Santiago, Chile\\
$^6$ Dipartimento di Astronomia, Universit\`a di Padova, Vicolo dell'Osservatorio 2, 35122, Padova, Italy\\
$^7$ Institute for Astronomy, University of Hawaii, 2680 Woodlawn Drive, Honolulu, HI, USA\\
$^8$ Department of Physics and Astronomy, Michigan State University, East Lansing, MI 48824-2320, USA \\
$^9$ Macalester College, 1600 Grand Avenue, Saint Paul, MN 55105, USA \\
$^{10}$ Cerro Tololo Inter-American Observatory, Casilla 603, La Serena, Chile\\
}    
}
\begin{abstract} 
We present the first results of a  variable star search in the field and in the globular clusters 
of the Fornax dwarf spheroidal galaxy. Variable stars were identified using the Image Subtraction Technique \citep{alard00} 
on time-series data obtained with the ESO 2.2 m and the Magellan 6.5 m telescopes. The variable star sample includes RR Lyrae stars,
Dwarf Cepheids and Anomalous Cepheids. The pulsation properties (namely: periods, light curves, period-amplitude relations 
and classification in Oosterhoff types) of Fornax variables from the present study are
discussed in some detail.
\end{abstract}
\section{Why to study the variable stars in an external galaxy?}                
Pulsating variable stars such as RR Lyrae stars, Dwarf Cepheids and Anomalous Cepheids are very interesting astronomical
objects that can be used to derive important information on the stellar system they belong to. They can be 
easily identified, even in high crowding conditions, thanks to their light variation, and their periods are 
independent from reddening and distance. Variable
stars are good tracers of the different stellar generations in a galaxy; in particular the RR Lyrae stars trace 
the oldest population,
and the Anomalous Cepheids the intermediate age populations, allowing  one to identify different star formation
episodes that have occurred in the host stellar systems. Variable
stars are also \emph{standard candles} and can be used to measure distances: the RR Lyrae stars through
 their nearly constant absolute magnitude in the {\it V} band, and the
Cepheids because they obey a \emph{period-luminosity relation}. 
Being among the oldest stars in a galaxy the RR Lyrae stars witnessed the first epochs of the galaxy formation, 
thus they provide information on the galaxy formation mechanism and assembling (\citeauthor{c04} \citeyear{c04}; \citeauthor{clementini04} \citeyear{clementini04}) .

We have made a comprehensive and deep study of the short period variable stars in the Fornax 
dwarf spheroidal galaxy (dSph) mapping its classical instability 
strip from the Dwarf Cepheids ({\it V} $\sim$ 24$-$25 mag)
  up to the Anomalous Cepheids ({\it V} $\sim$ 19 mag) in the field and in the globular cluster (GC) system of the galaxy, using wide field imaging obtained with the mosaics 
of the 2.2 m ESO-MPI  and the 4 m Blanco CTIO telescopes, 
and deep, high spatial resolution photometry of the clusters taken with the 6.5 m Magellan/Clay 
telescope.
Here, we report preliminary results from the analysis of the data covering a northern portion of the Fornax galaxy, along
with results on its clusters 2, 3, 4, and 5.

\section{Fornax dSph}                      
Fornax is a dwarf spheroidal galaxy placed at about 140 kpc from the Milky Way. 
The galaxy color-magnitude diagram (CMD) shows the presence of an old, an 
intermediate and a young stellar population (\citeauthor{buonanno85} \citeyear{buonanno85}, \citeauthor*{stetson98} \citeyear{stetson98}, \citeauthor*{saviane00} \citeyear{saviane00}). 
Fornax, Sagittarius and, possibly, the newly discovered Canis Major, are the only dwarf spheroidal galaxies known to contain globular clusters.
 The Fornax GC system is formed by five clusters (\citeauthor{bh39} \citeyear{bh39}; \citeauthor{h61} \citeyear{h61}, \citeyear{h65}, \citeyear{h69}). 
Previous photometric studies of Fornax GCs show that their HB morphologies are consistent with 
the presence of RR Lyrae stars (\citeauthor{buonanno98} \citeyear{buonanno98}, \citeyear{buonanno99}). 
This offers a great opportunity to investigate the variable star 
populations in extragalactic globular clusters, and to compare their properties with those of the field variables in Fornax and 
in the Milky Way field and clusters.

One of the most intriguing properties of the variable stars in  the Galactic GCs is the Oosterhoff 
dichotomy (\citeauthor{oo39}, \citeyear{oo39}): a sharp subdivision into two distinct types,
named Oosterhoff type I (OoI) and II (OoII), respectively, according
to the mean periods of the \emph{ab}-type RR Lyrae stars (OoI: $\langle$Pab$\rangle$= 0.55 d, OoI:
$\langle$Pab$\rangle$= 0.65 d). Do extragalactic globular clusters such as those in Fornax and the galaxy 
field variables conform to the Oosterhoff dichotomy observed in the Milky Way?

\citeauthor{bw02} (\citeyear{bw02}) made the first search for variables in the field of Fornax and identified  
515 candidate RR Lyrae stars, on an area of about 0.5 deg$^2$ covering the central region of 
the galaxy.
They concluded that the field RR Lyrae stars in Fornax have properties intermediate
between the two Oosterhoff types. 
However, since the magnitude of the RR Lyrae stars is close to the detection limit of their photometry, and 
because their data are not good enough to determine the type of variable on the basis of the light curve 
their analysis needs to be confirmed by better photometric data.

\citeauthor{mg03} (\citeyear{mg03}, hereafter MG03) identified candidate RR Lyrae stars in four of the
Fornax globular clusters
(namely Fornax 1, 2, 3 and 5) using HST archive data. Due to  
the short observational baseline and the small number of frames their data were not adequate for
a period search. Therefore, they determined periodicities by
fitting template RR Lyrae light curves to their data, and found that Fornax clusters possess mean characteristics 
intermediate between the two Oosterhoff
groups. 
Again, these results need to be confirmed on the basis of accurate and well sampled 
light curves allowing independent 
estimates of the periods. Our survey also allowed 
the first variability study of Fornax 4, the cluster in the central region of Fornax dSph.
    
\section{Data acquisition and reduction strategies}         
\noindent
We collected $B,V$ 
time-series data of Fornax with three different telescopes and instruments, namely the 
wide field imager of the 2.2 m ESO-MPI telescope,
the mosaic of the 4 m Blanco CTIO telescope,
and the Magic camera of the 6.5 m Magellan/Clay telescope.
Observations span the
time interval from November 2001 to December 2004. 
Central coordinates and field of view (FOV) of our pointings of Fornax dSph 
are summarized in Table~\ref{tab:osservazioni}, 
along with the number of frames per photometric band obtained at each 
pointing.  
\begin{table}
\caption{Photometric data of Fornax dSph.}
\begin{center}
\vspace{6pt}
\renewcommand{\arraystretch}{1.7}
\setlength\tabcolsep{6pt}
\tiny
\begin{tabular}{lclllcc}
\hline\noalign{\smallskip}
Telescope \& Instrum. &FOV& ~~~~Target &~~~$\alpha _{2000}$  & ~~~~$\delta _{2000}$  &N$_V$& N$_B$\\
\noalign{\smallskip}
\hline
\noalign{\smallskip}
2.2 m ESO WFI & 33$^{\prime}$ $\times$ 34$^{\prime}$ &Field\_1(For 3)  &$02^h 39^m 59^s$&$-34^\circ 10^\prime 00^{\prime\prime}$& ~17 & ~61\\
2.2 m ESO WFI & 33$^{\prime}$ $\times$ 34$^{\prime}$ &Field\_2(For 2,4)&$02^h 39^m 45^s$&$-34^\circ 39^\prime 00^{\prime\prime}$& ~16 & ~59  \\
4 m ~CTIO WFI & 36$^{\prime}$ $\times$ 36$^{\prime}$ &Field\_A(For 4,5)&$02^h 41^m 05^s$&$-34^\circ 17^\prime 30^{\prime\prime}$& 145&  ~64 \\
4 m ~CTIO WFI & 36$^{\prime}$ $\times$ 36$^{\prime}$ &Field\_B(For 2,3,4)&$02^h 39^m 24^s$&$-34^\circ 33^\prime 16^{\prime\prime}$ &~ 16&  ~8 \\
\hline
6.5 m Magellan/Clay  & 2.4$^{\prime}$ $\times$ 2.4$^{\prime}$  &For 2&$2^h 38^m 44.2^s$&$-34^\circ 48^\prime 33.1^{\prime\prime}$&~18&~6 \\
6.5 m Magellan/Clay  & 2.4$^{\prime}$ $\times$ 2.4$^{\prime}$  &For 3&$2^h 39^m 46.7^s$&$-34^\circ 15^\prime 23.1^{\prime\prime}$&~5&~5 \\
6.5 m Magellan/Clay  & 2.4$^{\prime}$ $\times$ 2.4$^{\prime}$  &For 4&$2^h 40^m 07.7^s$&$-34^\circ 32^\prime 16.2^{\prime\prime}$&58& ~20\\
6.5 m Magellan/Clay  & 2.4$^{\prime}$ $\times$ 2.4$^{\prime}$  &For 5&$2^h 42^m 21.3^s$&$-34^\circ 06^\prime 05.2^{\prime\prime}$&56& ~20 \\
\hline			
\end{tabular}		
\end{center}	       
\label{tab:osservazioni}							       
\end{table}
Photometric reductions have been completed for Field\_1 and for all the Magellan fields.
They were carried out with the packages DAOPHOT-ALLSTAR II \citep{st96} and ALLFRAME 
\citep{st94}. Variable stars were identified using the Image Subtraction Technique as performed within 
the package ISIS2.1 \citep{alard00}. This  
method is demonstrated to be very effective and allowed us to 
detect large numbers of bright variables as well as low-amplitude, faint variable stars 
such as 
the Dwarf Cepheids (DCs). These variables are about two magnitudes
fainter than the RR Lyrae stars. So far, our sample of DCs in Fornax 
is the largest one in an extragalactic stellar system.

Here we present preliminary results 
from the analysis of the variable stars detected in the four lower chips of the mosaic
of eight CCDs covering Fornax Field\_1 (Section 4), and in four Fornax globular clusters (Section 5).

\begin{figure}[hpbt]
\label{fig:field} 
\begin{center}
\caption{Color-magnitude diagrams of the four lower  CCDs covering Fornax Field\_1.
Confirmed pulsating variable stars in Chip 6 and 7 (upper two panels of the figure) are marked by different filled symbols: Dwarf Cepheids (filled triangles, 
in yellow in the electronic edition of the paper); 
RR Lyrae stars (filled circles, red: single mode pulsators, blue: double mode pulsators); and  
Anomalous Cepheids (filled squares, green). Filled circles in the two lower panels of the figure are
candidate variable stars detected in Chip 5 and 8, which still need to be confirmed and classified.}
\end{center}
\end{figure}

\section{Fornax Field\_1}         

Fornax Field\_1 covers 
 a 33$^{\prime} \times 34^{\prime}$ area 
North of Fornax dSph center, with the regions of higher stellar density found in Chip 6 and 7 
of the mosaic, and cluster Fornax 3 also falling in Chip 6.
Figure 1 shows the CMDs of the lower four CCDs of Fornax Field\_1, which 
cover an area of about 0.16 deg$^2$, and are found to contain a total number of 
706 candidate variable stars (displayed as large filled symbols). 
\begin{table}
\caption{Candidate variables in the lower four CCDs of Fornax Field\_1.}
\begin{center}
\vspace{6pt}
\renewcommand{\arraystretch}{1.4}
\setlength\tabcolsep{3pt}
\begin{tabular}{lccccl}
\hline\noalign{\smallskip}
 CCD & RR Lyrae stars& ACs & DCs&Binaries&Total\\
 \hline
 Chip 6 & 110  &  11  &  23  &   15& ~159 \\
 Chip 7 & 137  & $-$  &  23  &  15& ~175 \\
 Chip 5 & $-$  & $-$  &  ~7  &  $-$& ~209$^{\mathrm{a}}$ \\
 Chip 8 & $-$  & $-$  &  14  &  $-$& ~163$^{\mathrm{a}}$ \\
\hline			
\end{tabular}	
\label{tab:somme}	
\end{center}
$^{\mathrm{a}}$ \scriptsize{The study of the light curves and the classification in types for the variables in 
Chip 5 and 8 is still in progress, only the DCs have been classified so far. Total numbers 
for these two chips may include spurious detections.}
\label{tab:Tab1a}							       
\end{table}
Study of the light curves,  definition of the period, and classification in types have been 
completed only for the variable stars in Chip 6 and 7. These confirmed variables are marked by  
different symbols in the two upper panels of Figure 1, namely:
filled triangles are Dwarf Cepheids (DCs); filled circles are RR Lyrae stars; 
filled squares are Anomalous Cepheids (ACs).
The filled circles in the two lower panels of Figure 1 are candidate variables that still
need to be confirmed and classified; however, their position  on the CMD  is consistent 
with many of them being RR Lyrae stars and Dwarf Cepheids.
Variable stars in the upper left panel of Figure 1 are plotted according 
to their intensity-averaged magnitudes and colors. 
They define very well the region of the classical instability strip in Fornax dSph.
Candidate variables in the other 
panels of the figure span a much larger magnitude/color range because they are plotted at random phase. 
The number of candidate variables in each of the four CCDs analyzed so far and their classification in types, when available,
are summarized in Table~\ref{tab:somme}.
\begin{table}[hpbt]
\caption{Variable stars in Chip 6, divided by field and cluster For 3.}
\begin{center}
\vspace{6pt}
\renewcommand{\arraystretch}{1.4}
\setlength\tabcolsep{3pt}
\begin{tabular}{lccccc}
\hline\noalign{\smallskip}
 Target & RRab,c & RRd & ACs&DCs&Binaries\\
 \hline
 ~Field  & 72  &10  &  9  &  21 &15  \\
 ~For 3 & 18 & 10 & 2  & 2 & - \\
\hline			
\end{tabular}		
\end{center}	       
\label{tab:Tab1b}							       
\end{table}

We have already confirmed and classified in types 355 of the variable stars in Table \ref{tab:somme}. 
This number gives a lower limit for the variable star density in Fornax twice as much as that derived by 
\citeauthor{bw02} (\citeyear{bw02}). In Table~\ref{tab:Tab1b} we summarize results 
for Chip 6 dividing the variable stars between field and
cluster.
\begin{figure}[hpbt]
\begin{center}
\includegraphics[scale = 0.45]{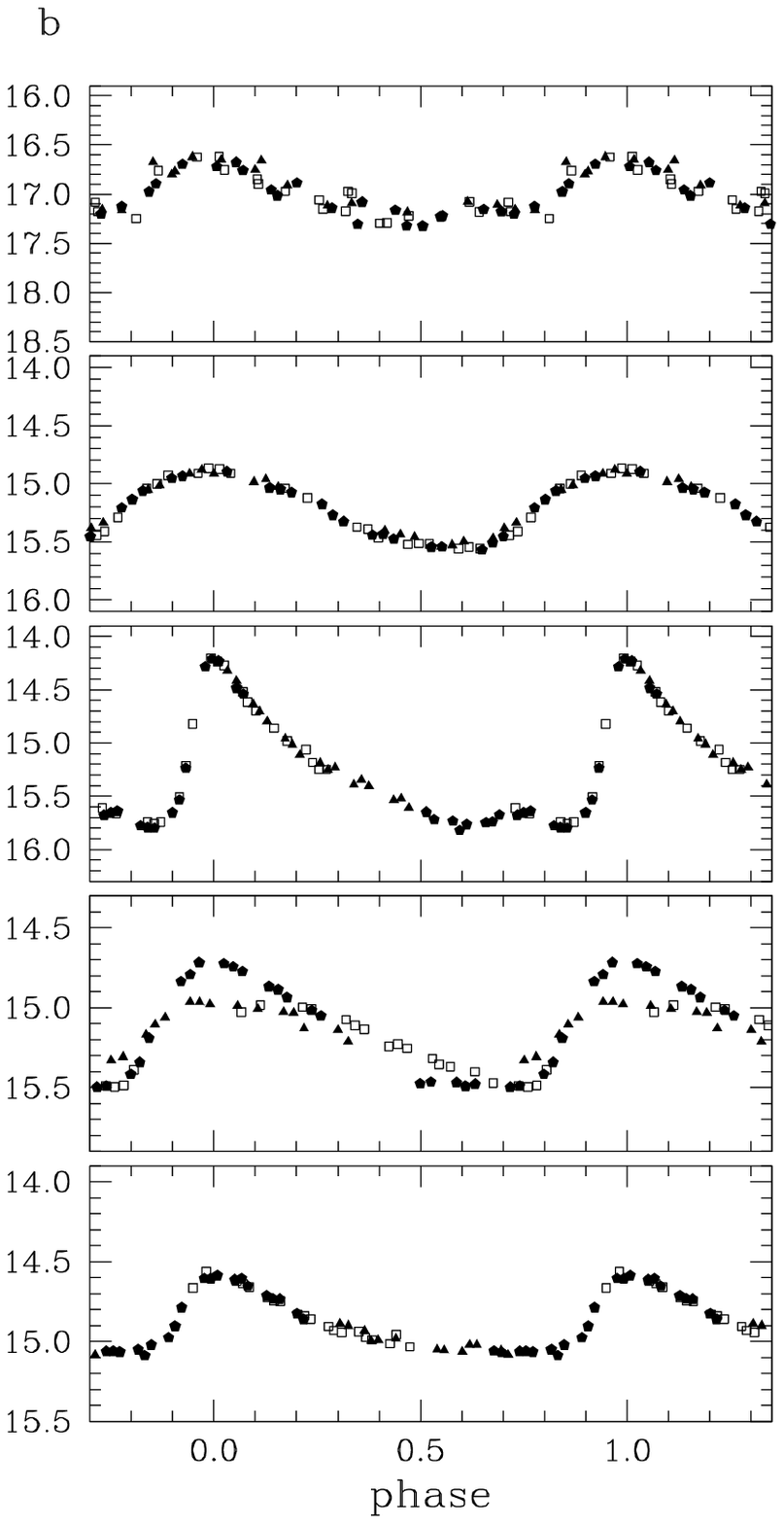}
\includegraphics[scale = 0.45]{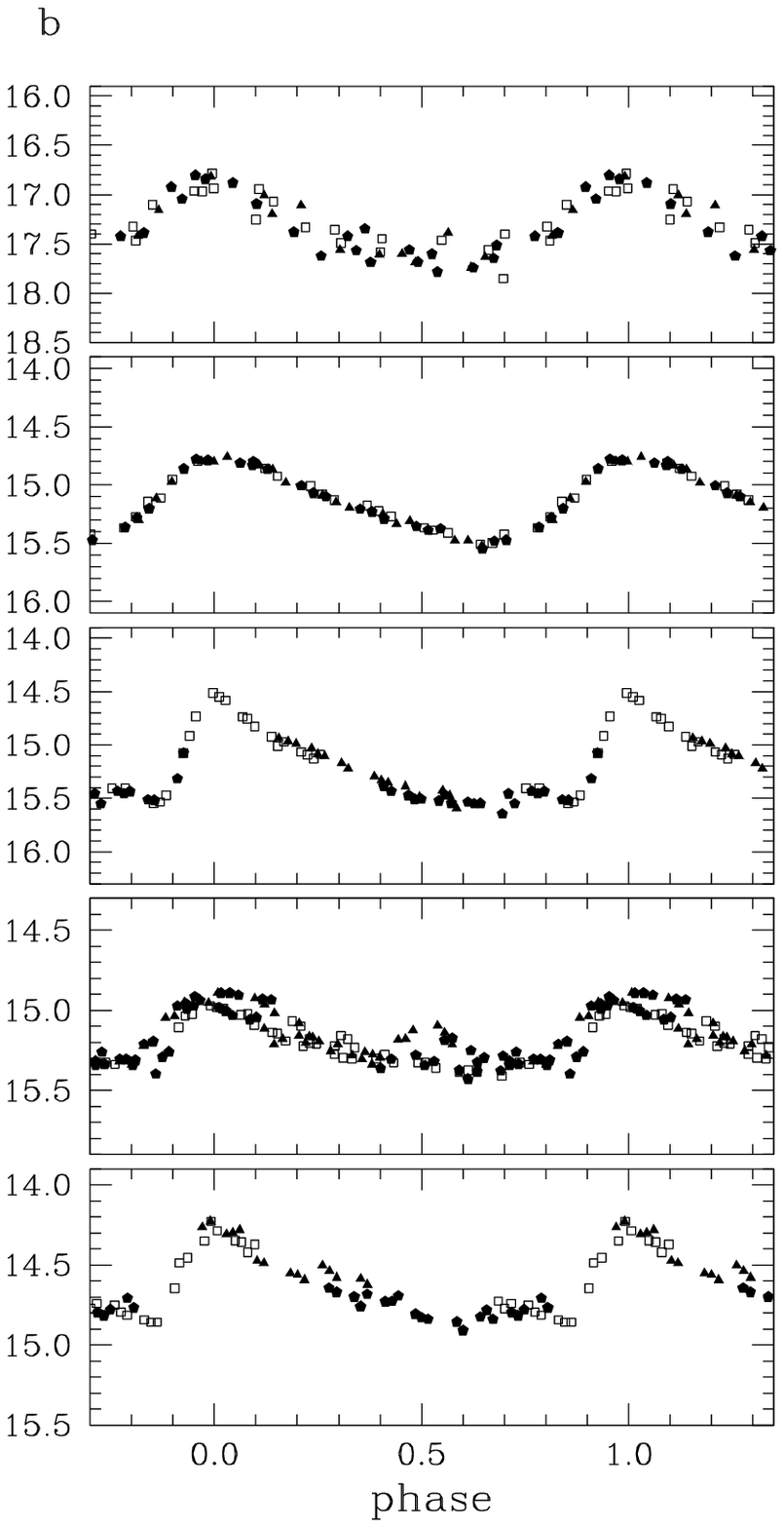}
\caption{Instrumental $B$ light-curves  for
variable stars in Chip 6. Left panels: field variables; right panels: variable stars
in For 3. From top to bottom: DCs, {\it c-}, {\it ab-}, and {\it d-}type RR Lyrae stars, ACs.}
\label{fig:lc} 
\end{center}
\end{figure}
Figure~\ref{fig:lc} shows examples of light curves for the field (left panels) and
cluster (right panels) variable stars in this CCD. Variable stars in Chip 6 have been fully analyzed, and in the left panel of 
Figure~\ref{fig:plpa} we plot their period-luminosity diagram in the $V$ band. 
We can see how both Dwarf and Anomalous Cepheids follow a \emph{period-luminosity relation}, 
while the RR Lyrae stars show a nearly constant average magnitude.
The mean period of the field RR Lyrae stars in Chip 6 is 0.595 d (r.m.s.= 0.039), for the 
{\it ab-}type, and 0.361 d (r.m.s.= 0.040) for the {\it c-}type variables, respectively. 
These values confirm the intermediate Oosterhoff type of Fornax field RR Lyrae stars
(see also right panel of Figure~\ref{fig:plpa}). We used the mean period of the {\it ab-}type
 RR Lyrae stars to estimate the 
metallicity of the old stellar population in Fornax. Adopting \citeauthor{sandage03}'s (\citeyear{sandage03}) relation 
 $\log \langle P_{ab} \rangle  =-0.092 \left[{\rm Fe/H}\right] -0.0389$ we obtain: 
$\left[{\rm Fe/H}\right] _{\rm field}= -1.78$.
The average $V$ magnitude of the field RR Lyrae stars in Chip 6 is:
$\langle${\it V(RR)}$_{field}$ $\rangle=21.281 \pm 0.100$ mag. This value leads to a distance modulus of:
$\mu_{\rm Fornax}$=20.72$\pm 0.10 $ mag, (on the assumption of:
$M_V(RR) =0.50$ mag at $\left[{\rm Fe/H}\right] =-1.5$, \citeauthor{g03} \citeyear{g03}; 
$\Delta M_V(RR) /  \left[{\rm Fe/H}\right]  = 0.22 \ \mbox{mag/dex}$, \citeauthor{g04} \citeyear{g04};  
$E(B-V)=0.04 \pm 0.03$, and the standard extinction law), in
very good agreement with estimates by \citeauthor{buonanno99} (\citeyear{buonanno99}), \citeauthor{saviane00} (\citeyear{saviane00}), and MG03.
\begin{figure}[hpbt]
\plottwo{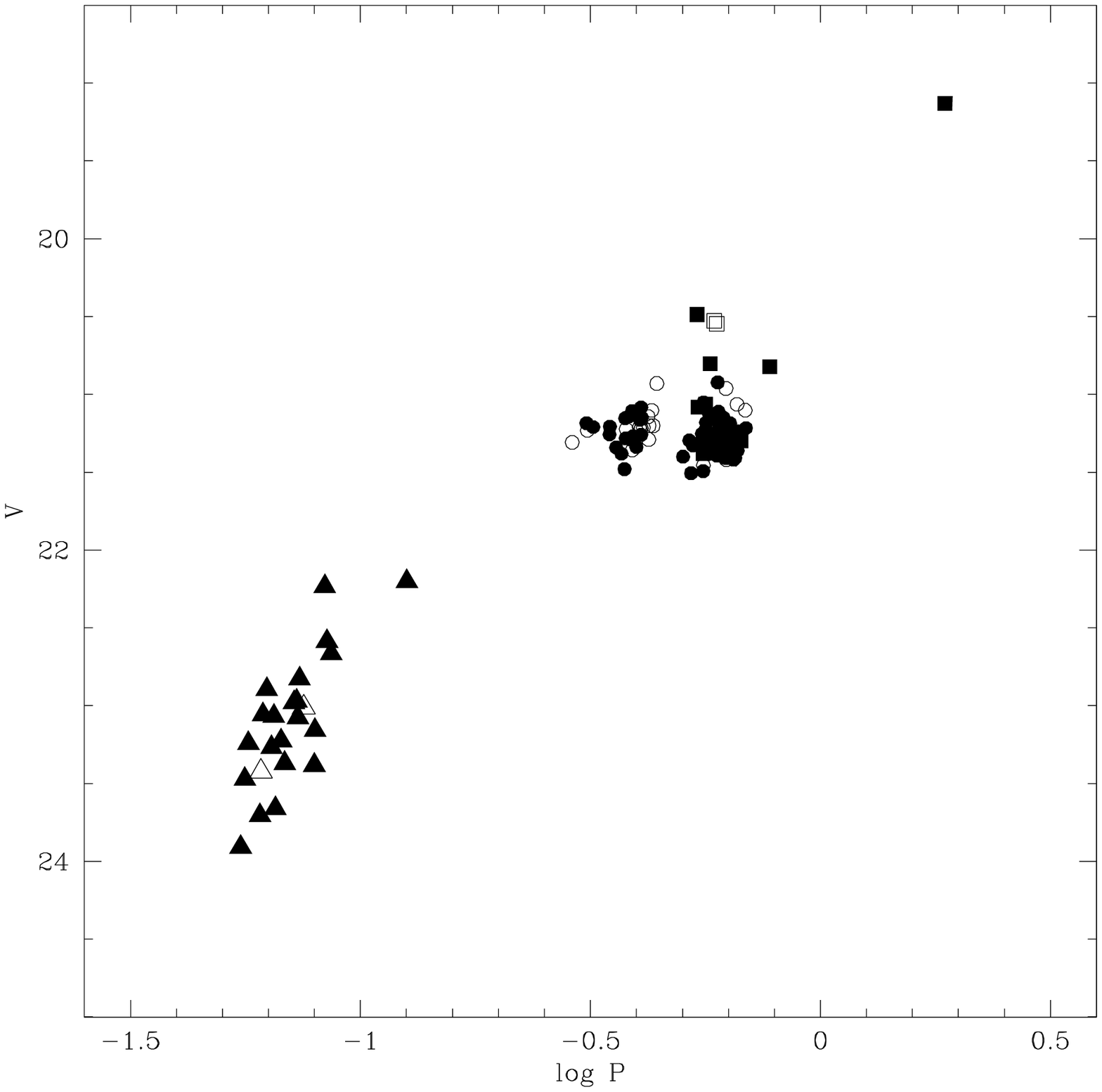}{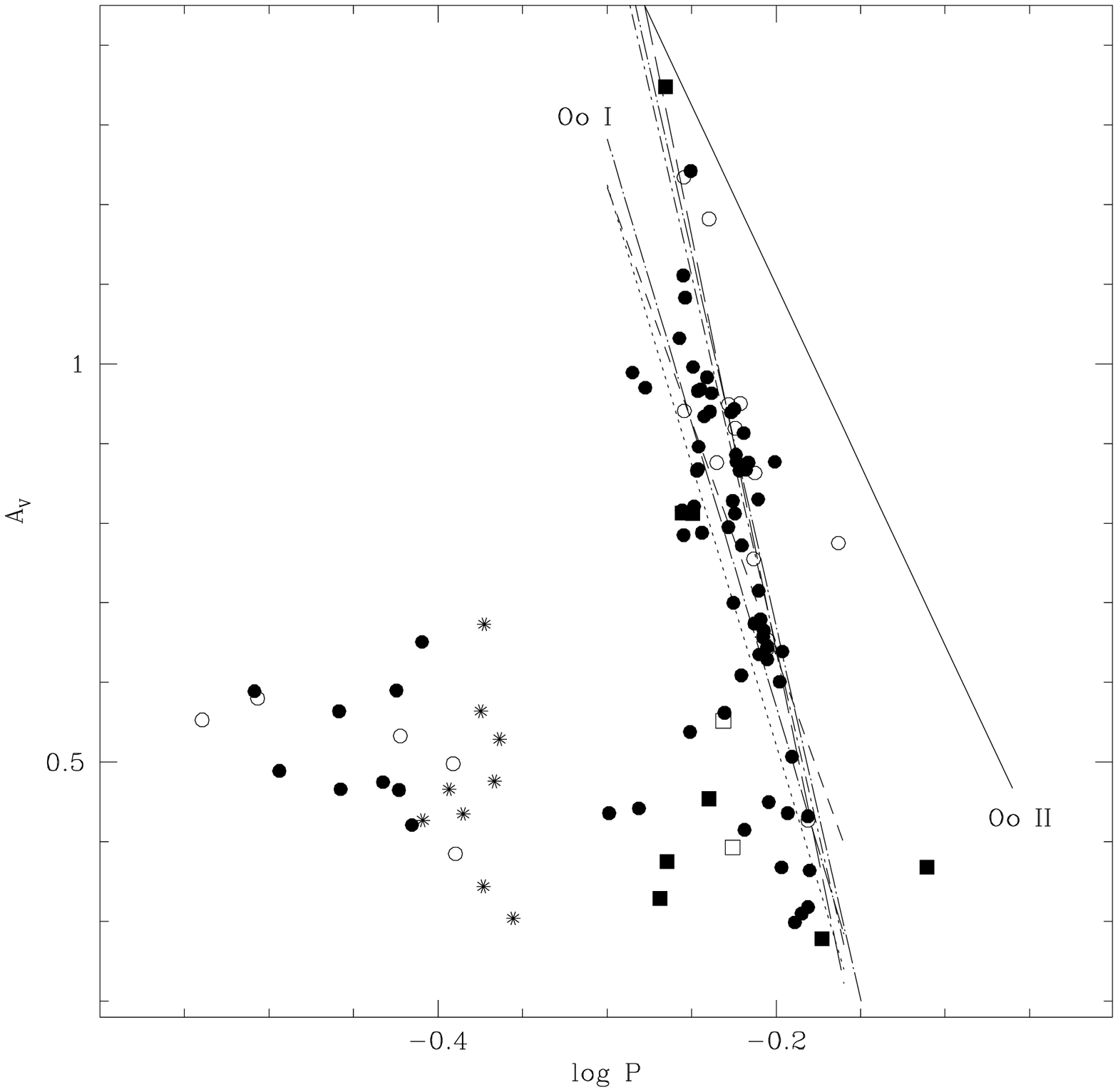}
\caption{Period-Luminosity (on  the left) and Period-Amplitude (on the right) 
relations in the $V$ band for variable stars
in Chip 6. Filled symbols are field variables, open symbols are variable stars in For 3.
Triangles are DCS, circles are RR Lyrae stars, squares are ACs. Lines in the right panel
are the Period-Amplitude relations for OoI and OoII Galactic GCs from \citeauthor{cr00} (\citeyear{cr00}; solid lines) and for some dwarf spheroidal galaxies 
: And VI (dotted line); Sculptor (short-dashed line); Leo II (long-dashed line); Draco (dotted-dashed lines) from \citeauthor{pr02}(\citeyear{pr02});
and Leo I (dotted-long-dashed line) 
from \citeauthor{h01}(\citeyear{h01}).}
\label{fig:plpa}
\end{figure}	

\section{Variable stars in Fornax globular clusters}         

\begin{figure}[hpbt]

\begin{center}
\caption{Color-magnitude diagrams of Fornax GCs 3,4,5, and 2 from the ESO-WFI data (For 3) and the Magellan/Clay observations (For 2, 4 and 5). Variable stars are
marked by solid symbols, as in Figure 1.}
\label{fig:cmfg}
\end{center}
\end{figure}
Well sampled light 
curves were obtained for the variables in Fornax 3, 5 and 4. Variable stars were identified also in Fornax 2, however
data for this cluster do not
allow a reliable definition of the periods.
Results obtained for the four clusters are summarized in Table~\ref{tab:resultgc}.
\begin{table}[hpbt]
\caption{Results on the variable stars in Fornax clusters 2, 3, 4, and 5.}
\begin{center}
\vspace{6pt}
\tiny
\renewcommand{\arraystretch}{1.4}
\setlength\tabcolsep{6pt}
\small
\begin{tabular}{clccccc}
\hline\noalign{\smallskip}
 Cluster & RR Lyrae stars & DCs & ACs& $\langle$ P$_{ab}$ $\rangle $ & $\langle $ P$_{c}$ $\rangle $&~~$\left[{\rm Fe/H}\right]$$^{\mathrm{a}}$\\
 \hline
 For 3   & 28(13ab,5c,10d)       &2    &  2 &  0.606       &0.358       & $-$1.96  \\
 For 4   & 18(14ab,3c,1d)       & -   &  - &  0.592      &0.359       & $-$2.01 \\
 For 5   & 17(9ab,7c,1d)       & 1   & -  & 0.576        &0.353       & $-$2.20 \\
 For 2   & 10       & -   & -  & -             & -         & $-$1.79\\
\hline			
\end{tabular}		
     \end{center}
$^{\mathrm{a}}$\scriptsize{Metal abundances are from Buonanno et al. (1998,1999)}
\label{tab:resultgc}							       
\end{table}
The average periods of the {\it ab-}type RR Lyrae stars in For 3 and 5
are in good agreement with the values derived by MG03 for their subsamples of RRab's with good periods.
Our $\langle $ P$_{c}$ $\rangle $ values are instead shorter than in MG03. This is particularly true of For 3, the cluster where we
also detected a large population of double mode pulsators, twice in number the cluster RRc variables. This
type was not recognized by MG03 and may have contaminated their RRc sample. These authors also found in For 3 several very 
long-period RRc's, that we did not find in our period search.
\begin{figure}[hpbt]
\begin{center}
\includegraphics[scale = 0.45]{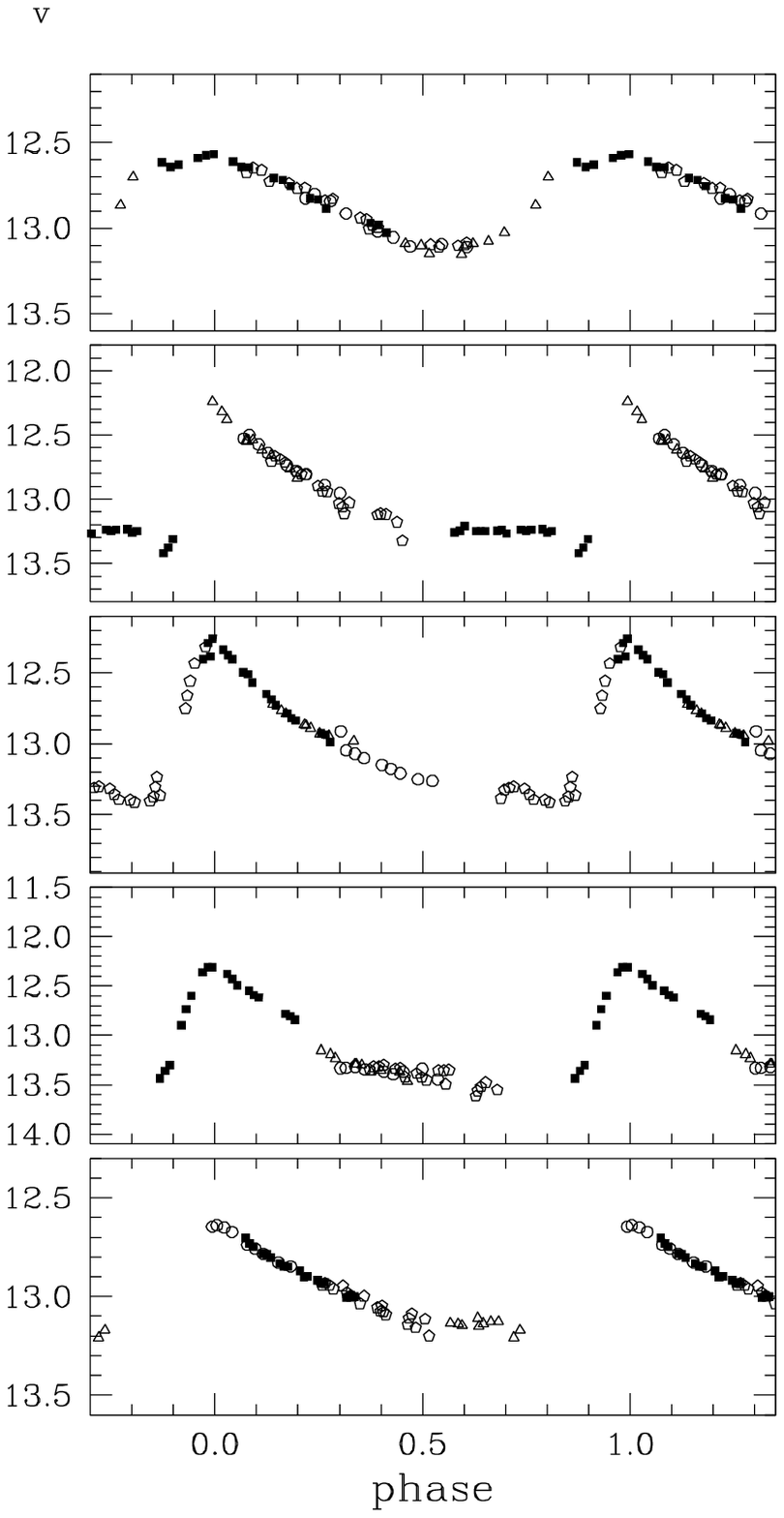}
\includegraphics[scale = 0.45]{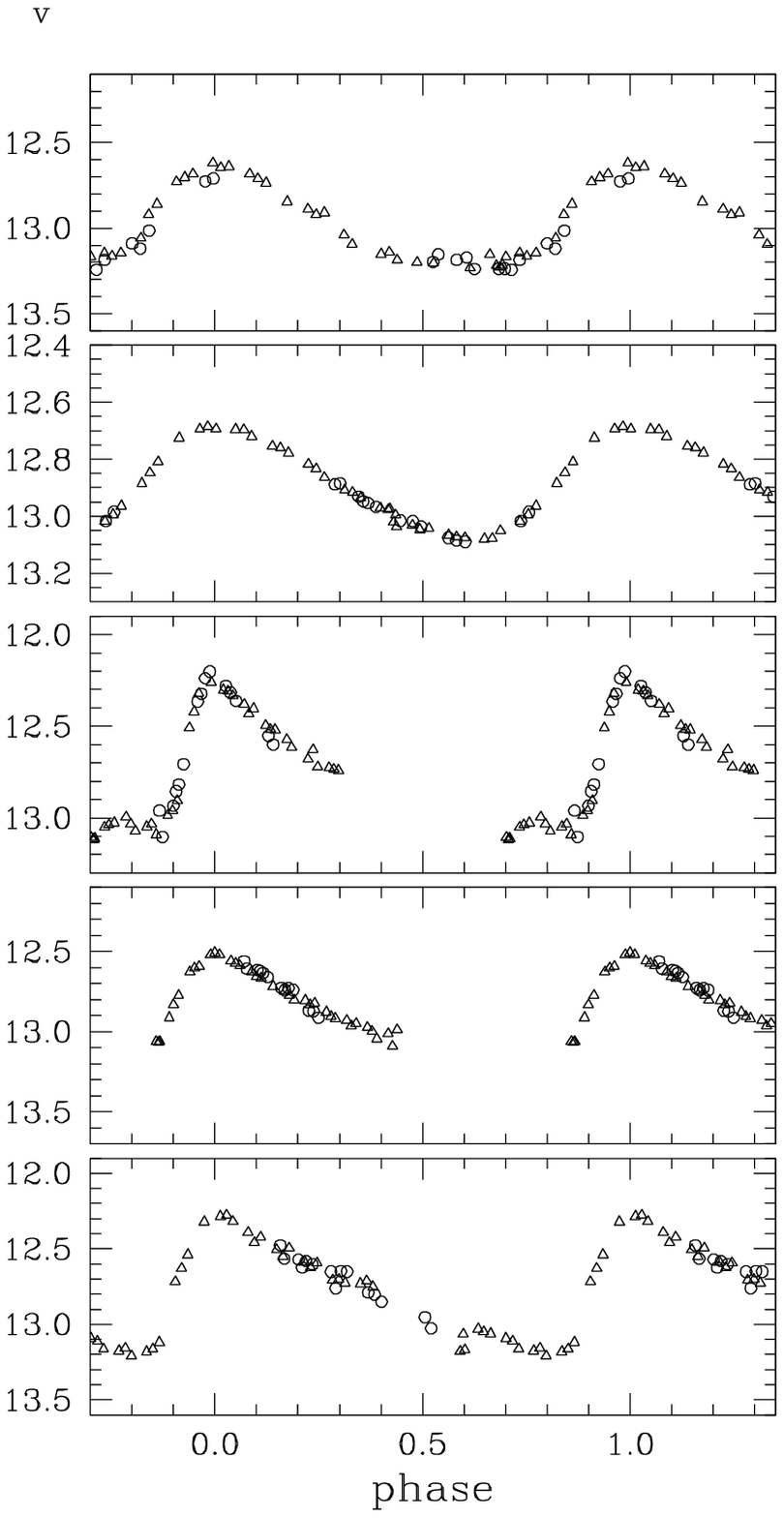}
\caption{Instrumental $V$ light-curves  for {\it ab-} and {\it c-}type
RR Lyrae stars in Fornax clusters 4 (left panels) and 5 (right panels).}
\label{fig:lcfgc} 
\end{center}
\end{figure}

\begin{figure}[hpbt]
\begin{center}
\plotone{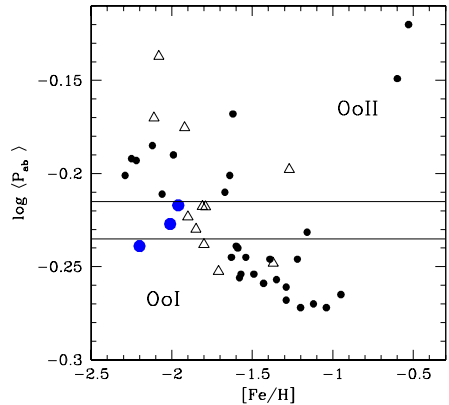}
\caption{Position occupied in the Oosterhoff plane by the 
Galactic GCs (dots; from \citeauthor*{castellani03} \citeyear{castellani03}), the Large Magellanic Cloud clusters (triangles), and
Fornax clusters For 3, 4 and 5 (big circles, in blue in the electronic edition
of the journal; from the present study).} 
\label{fig:oo} 
\end{center}
\end{figure}
Figure~\ref{fig:cmfg} shows the CMDs of the four clusters with the variable stars
marked by filled symbols. Examples of light curves of {\it ab-} and {\it c-}type RR Lyrae stars
in For 4 and 5 are shown in Figure~\ref{fig:lcfgc}. Using the average periods of the {\it ab-} type RR Lyrae stars derived in our study
for the variables in For 3, 4 and 5 we can now check the location of 
 Fornax GCs in the 
Oosterhoff diagram, showing the relationship between the average period of {\it ab-}type RR Lyrae stars and the metallicity. This is shown in Figure~\ref{fig:oo}.
We find that Fornax 3 falls slightly below the edge of the Oosterhoff II region, while Fornax 5
is at the edge of the OoI region. 
Fornax 4 with $\langle$ $P_{ab}$ $\rangle $=0.59 d 
falls inside the ``Oosterhoff gap''. However, this cluster shows 
two separate peaks in the period distribution of the {\it ab-}type RR Lyrae stars,
with average values respectively around 0.55 and 0.65 d, as if, similarly to NGC1835 in the LMC (\citeauthor{so03}\citeyear{so03}), there is an Oosterhoff 
dichotomy within the cluster itself.




\end{document}